\documentclass[12pt]{iopart}

\usepackage{graphicx}

\newcommand{\CuC}{(Cu,C)Ba$_{2}$Ca$_{2}$Cu$_3$O$_{9\pm\delta}$}
\newcommand{\cuc}{(Cu,C)Ba$_{2}$Ca$_{3}$Cu$_4$O$_{11\pm\delta}$}
\newcommand{\CuCTarget}{Ba$_{2}$Ca$_{3}$Cu$_{4.6}$O$_{y}$}

\newcommand{\REBCO}{REBa$_2$Cu$_3$O$_{7-\delta}$}

\begin{document}

\title{Growth of (Cu,C)Ba$_{2}$Ca$_{2}$Cu$_3$O$_{9\pm\delta}$ thin films on flexible Hastelloy tapes} 

\author{Meng-Jun Ou, Yuecong Liu, Yi Wang, Hai-Hu Wen*}
\ead{hhwen@nju.edu.cn}
\address{National Laboratory of Solid State Microstructures
	and Department of Physics, Collaborative Innovation Center of Advanced Microstructures, Nanjing University, Nanjing 210093, People’s Republic of China}

\vspace{10pt}

\begin{abstract}
The applications of superconducting cable or magnet require that the superconductors are made into wires or tapes. For cuprate superconductors, this is a big challenge because of the strong flux motion induced by high anisotropy, very short coherence length and strong thermal fluctuation, etc.  One of the ways is to fabricate superconducting films on flexible metallic tapes with oxide buffer layers. The successful one so far is the REBa$_2$Cu$_3$O$_7$ (REBCO, RE=rare earth elements) films in tape form, as called the coated conductors. While the superconducting transition temperature of REBCO system is limited to about 90 K. Here we report the successful fabrication of another new non-toxic superconducting film, namely (Cu,C)Ba$_{2}$Ca$_{2}$Cu$_3$O$_{9\pm\delta}$ on these flexible metallic tapes with LaMnO$_3$ and CeO$_2$ as the top layers. The onset superconducting transition occurs at 112 K and 110 K, and the zero-resistance transition temperatures are about 96 K and 98 K, respectively. The temperature dependent resistivity under magnetic fields in different directions reveal a relatively small anisotropy. Further optimization of the films will improve the zero resistance transition temperature, thus can also improve the characteristic properties for applications. Our results show that the (Cu,C)Ba$_{2}$Ca$_{2}$Cu$_3$O$_{9\pm\delta}$ is a promising candidate material for the high power applications in liquid nitrogen temperature region.
\end{abstract}

%
\noindent{\it Keywords}: (Cu,C)-1223, Application, Thin films

\section{Introduction}

Since the discovery of superconductivity\cite{Onnes1911}, people immediately realized that superconductors may be applied in many ways, thus enormous efforts have been dedicated to explore practical superconductors that meet the industrial requirements. Superconductors can be categorized into low-temperature superconductors (LTS) and high-temperature superconductors (HTS) according to their critical temperatures. Applicable low-temperature superconductors include NbTi, Nb$_3$Sn and so on. NbTi superconductors have very good ductility and can be directly made into long wires\cite{Zhang2019}. Nb$_3$Sn, on the other hand, due to its brittleness and hardness, is currently prepared by the bronze process and powder-in-tube (PIT) method\cite{Xu2017}. Both superconductors are currently commercialized. The critical temperature of some cuprate high temperature superconductors is higher than the liquid nitrogen boiling temperature (77 K), such as Hg, Tl, Bi and RE-based systems\cite{SchillingN1993,ShengN1988,MaedaJJoAP1988, CavaNature1987}. Due to the existence of toxic elements and the high anisotropy, the application of Hg-based and Tl-based systems is limited. Bi-based system, as a ceramic material, can not be directly made into wires and are mainly manufactured by powder-in-tube method, known as the first generation of high-temperature superconducting tapes. Because of its strong anisotropy, the vortex motion is very strong leading to a rather low irreversibility magnetic field in the liquid nitrogen temperature region, and the critical current density also decreases quickly in a small magnetic field at 77 K\cite{Clayton2004}. The critical temperature of \REBCO\ (REBCO) is about 90 K, and the anisotropy is relatively small compared with the Bi-system\cite{AlNaafa2010}, which means that there is a high irreversibility field in the liquid nitrogen temperature region. However, due to the very small coherence length and the small tolerant misorientation angle between neighboring grains for the superconducting current to go through, the PIT technique was not successful to make the long wires for REBCO. In order to eliminate the weak links from the grain boundaries and obtain a better grain alignment to meet the requirements for practical applications, REBCO can be fabricated by epitaxial thin film method such as metal-organic chemical vapour deposition(MOCVD) and pulsed laser deposition (PLD)\cite{MacManusDriscoll2021}, etc. Thus the REBCO is known as the second generation of high-temperature superconducting tapes. However, the critical temperature of REBCO is about 90 K, lower than other systems with three or four layers of $CuO_2$ planes in the superconducting block. There is an another high-temperature superconducting system, (Cu,C)Ba$_{2}$Ca$_{n-1}$Cu$_{n}$O$_{2n+3}$ (n=1,2,3, ...) \cite{KawashimaPCS1994,KumakuraPCS1994,ZhangSA2018,HeSSaT2021,HeMTP2022}. Here \CuC ((Cu,C)-1223) and \cuc ((Cu,C)-1234)  have a superconducting transition temperature of about 120 K and a higher irreversibility field than REBCO. But (Cu,C)-1223 and (Cu,C)-1234 bulks can only be synthesized by high-pressure methods, strongly limiting their applications. The growth of (Cu,C)-based superconducting films with high transition temperature is still challenging. The transition temperatures of  (Cu,C)-1223 and (Cu,C)-1234 films grown on single-crystal substrates have reached 99.7 K and 96 K ~\cite{DuanSSaT2020, DuanPCSaiA2020,Ou2024}, respectively. Similar to REBCO, these materials need to be epitaxially grown on flexible metallic substrate to meet the demands of industrial applications. However, to the best of our knowledge, up to now, there are no reports of (Cu,C)-1223 films grown on flexible metallic substrates. Due to its superior properties, it is important to grow this material on flexible metallic substrate in order to meet the demand for the industrial applications.

In this work, we successfully fabricated c-axis oriented (Cu,C)-1223 thin films on flexible metallic substrates with the top layer of LaMnO$_3$ and CeO$_2$ by pulsed laser deposition technique, with zero-resistance transition temperature of 96 K and 98 K, respectively. The anisotropy estimated from the upper critical magnetic field is about 5.3, close to the value $\Gamma =5$ at 90 K obtain by the anisotropic Ginzburg-Landau theory. In addition, the irreversibility field is 4.2 T at 77 K. And the calculated critical current density $J_c$(4.2 K) reaches 4.2$\times$10$^6$ A/cm$^2$. The successful growth of (Cu,C)-1223 on flexible metallic substrates and its superior properties suggest that the material has great potential for applications in the liquid nitrogen temperature region.

\section{Experimental details}
\begin{figure*}
	\centering
	\includegraphics[width=\textwidth]{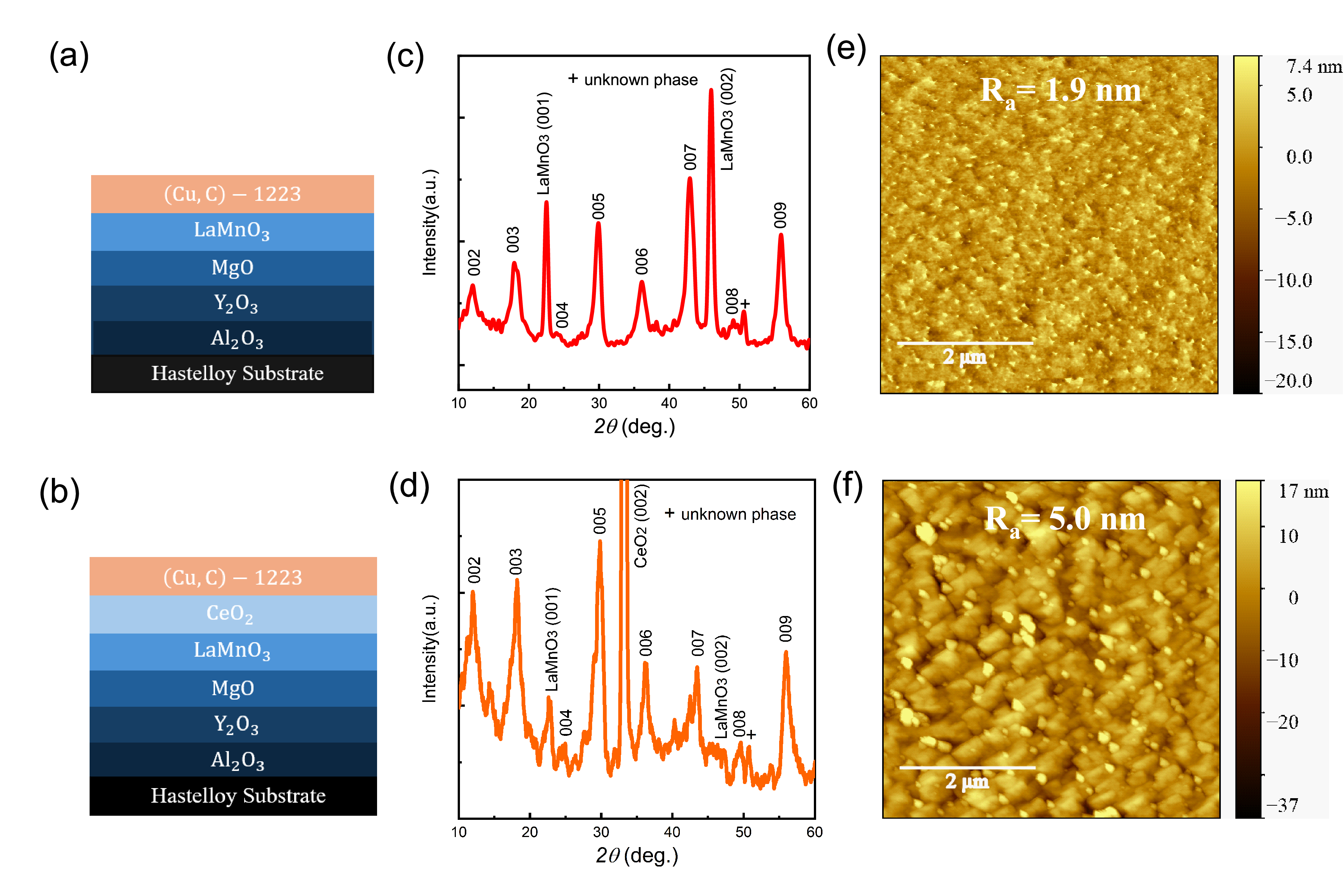}
	\caption{(a), (b) Sketch of two-type architectures of the (Cu,C)-1223 films. (c), (d) X-ray diffraction (XRD) patterns of the \CuC\ films on flexible Hastelloy substrates with the top layer of  LaMnO$_3$ and CeO$_2$, respectively. (e), (f) AFM images of the (Cu,C)-1223/LaMnO$_3$ film and the (Cu,C)-1223/CeO$_2$ film.}
	\label{fig1}
\end{figure*}

The \CuC\ thin films were epitaxially grown on Hastelloy tapes with LaMnO$_3$ and CeO$_2$ top buffer layer by pulsed laser deposition technique with a 248-nm KrF excimer laser, using the nominal composition target of \CuCTarget. The Hastelloy tapes with LaMnO$_3$ and CeO$_2$  top buffer layer are obtained from Companies of Suzhou Advanced Materials and Shanghai Superconductor, respectively. Fig.~\ref{fig1}(a) and (b) show the architectures of the (Cu,C)-1223 coated conductors on different top layers.  During deposition, a mixed gas of O$_2$ and CO$_2$ was introduced into the chamber, and the total pressure was maintained at 15-25 Pa. The deposition temperature for all films was controlled in the region 620-680 $\rm{^\circ C}$. The laser energy density of  1-2 J/cm$^2$ and repetition rate of 5 Hz were used to ablate the target. After the deposition, the films were cooled down to room temperature in deposition pressure at a rate of $5 \rm{^\circ C}/min$. 

The structural characterization and surface morphology information of the (Cu,C)-1223 thin films were obtained by x-ray diffractometer (XRD, Bruker D8 Advance) and atomic force microscopy (AFM, FSM-Precision). We measured the electrical resistance with magnetic fields from 0 to 9 T using the standard four probe method in a physical property measurement system (PPMS-9T, Quantum Design). The angle dependent resistivity was measured  with the angle $\theta$ varied from $0 \rm{^\circ}$ to $180 \rm{^\circ}$,  where $\theta=0 \rm{^\circ}$  represents the magnetic field $H\parallel c$. The current was always perpendicular to the external magnetic field and applied in the $ab$-plane. The magnetization measurements were performed in a SQUID VSM-7T (Quantum Design).

\section{Results and discussion}

The x-ray diffraction patterns of the (Cu,C)-1223 films on flexible Hastelloy substrates with LaMnO$_3$ and CeO$_2$ top buffer layer are displayed in Fig.~\ref{fig1}(c) and (d). All observed (00l) diffraction peaks can be indexed to (Cu,C)-1223 and the substrates, indicating that the films have been grown with c-axis orientation. Analysis of the XRD data yields a lattice constant $c$ = 14.82 \r A for both films. Surface morphology and roughness have been investigated through atomic force microscopy (AFM) as shown in Fig.~\ref{fig1}(e) and (f). The average roughness (R$_a$) of the (Cu,C)-1223/LaMnO$_3$ film is about 1.9 nm, determined over the scan area $5\times 5 \mu m^{2}$. For the (Cu,C)-1223/CeO$_2$ film, the average roughness is about 5.0 nm. One possible reason for the obvious different $R_a$ is that the growth of the (Cu,C)-1223 films (a=3.86 \r A) may be affected by epitaxial tensile strain from LaMnO$_3$ (a=3.94 \r A) and epitaxial compressive strain from CeO$_2$ (a=3.82 \r A).

\begin{figure*}
	\centering
	\includegraphics[width=\textwidth]{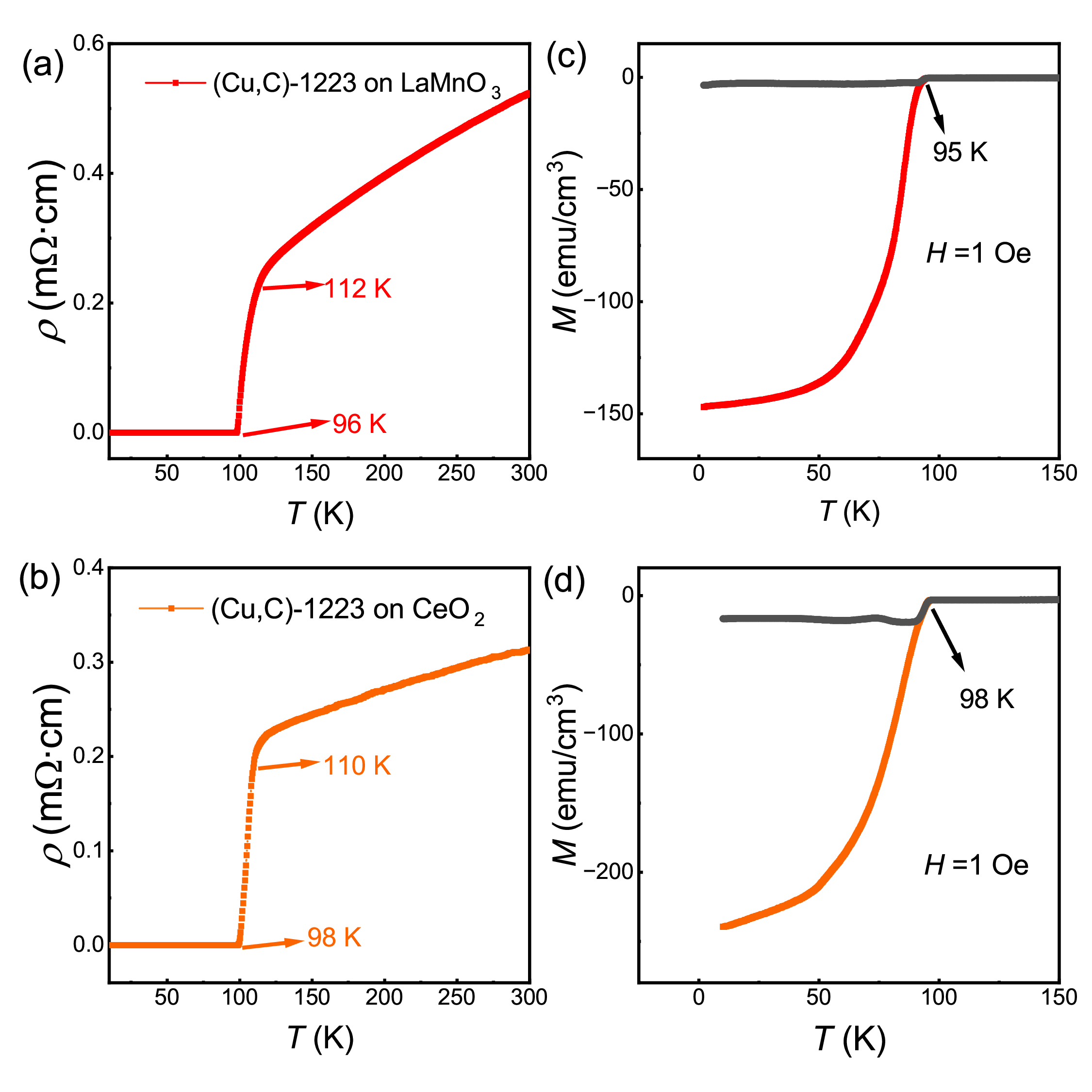}
	\caption{ (a), (b) Temperature dependent resistivity of the \CuC\ films on flexible metallic substrates with the top layer of  LaMnO$_3$ and CeO$_2$. (c), (d) The temperature dependence of magnetic susceptibility under a magnetic field of 1 Oe.}
	\label{fig2}
\end{figure*}
Fig.~\ref{fig2}(a) and (b) show the temperature dependence of the resistivity for the two kinds of films on different substrates. We can see that the resistivity displays metallic behavior in the normal state. For (Cu,C)-1223/LaMnO$_3$ film and (Cu,C)-1223/CeO$_2$ film, the onset superconducting transition temperatures ($T_c^{onset} $ defined by $90\% \rho_n$) are about 112 K and 110 K, and the zero-resistivity transition temperatures are about 96 K and 98 K, which are already higher than that of YBCO. If the deposition condition is optimized, the transition width would become narrower with high zero resistance transition temperatures. We performed magnetization measurement with the magnetic field (1 Oe) perpendicular to the film surface as shown in Fig.~\ref{fig2}(b). The superconducting transition occurs at 95 K and 98 K, quite close to the zero resistance transition temperatures. For (Cu,C)-1223 bulk samples, the c-axis lattice constant can change from 14.76 \r A to 14.82 \r A by reduction of oxygen, and at the same time the superconducting transition temperature increases from 67 K to 120 K~\cite{ChailloutPCS1996}. In addition, the transition temperature of the (Cu,C)-1234 film is affected by the flow rate of CO$_2$ ~\cite{DuanPCSaiA2020}. Thus the films in this work display higher transition temperature and larger c-axis lattice constant compared with the as-prepared bulk sample, possibly due to differences in oxygen and carbon content. Furthermore, the (Cu,C)-1223/LaMnO$_3$ film exhibits a smaller residual resistivity of 0.23 $\rho$(300 K) obtained by extrapolation of the resistivity down to 0 K. The x-ray diffraction peaks are sharper and the film surface are flat with the average roughness value of 1.9 nm. These results indicate that the (Cu,C)-1223 film on flexible metallic substrates with the top layer of LaMnO$_3$ has higher quality. Thus superconducting properties of the (Cu,C)-1223/LaMnO$_3$ film will be investigated in more detail in the following.
\begin{figure*}
	\centering
	\includegraphics[width=\textwidth]{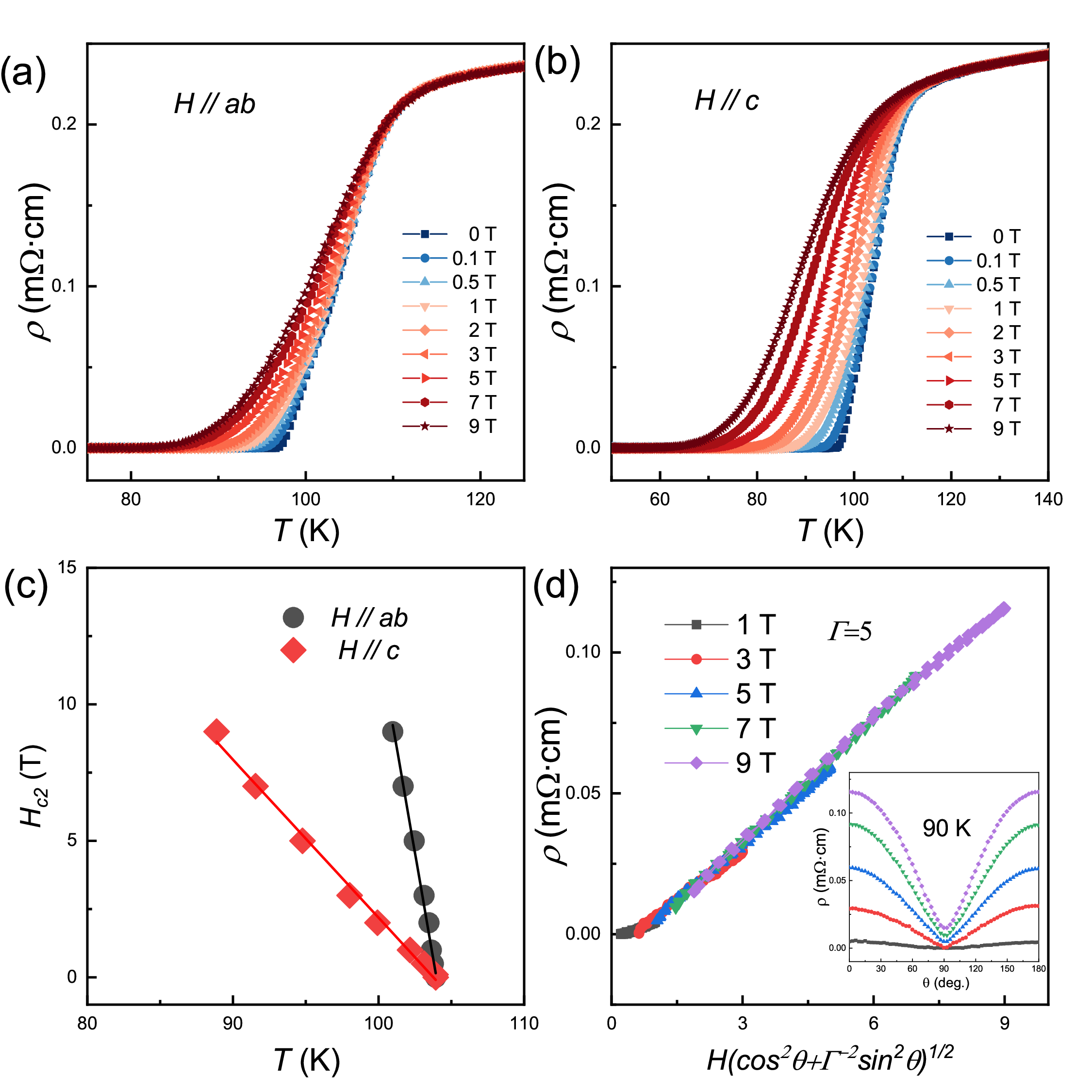}
	\caption{ (a), (b) Temperature dependent resistivity of the (Cu,C)-1223/LaMnO$_3$ film under magnetic fields $H\parallel ab$ and $H\parallel c$, respectively. (c)  Temperature dependence of the upper critical field $H_{c2}$. The solid lines show the linear fitting results. (d) The resistivity as a function of the effective magnetic field at 90 K. The inset shows that angle dependent resistivity under different magnetic fields.}
	\label{fig3}
\end{figure*}

The temperature-dependent magnetoresistance was measured under magnetic fields perpendicular ($H\parallel c$) and parallel ($H\parallel ab$) to the film surface, up to 9 T. Fig.~\ref{fig3}(a) and (b) show the temperature-dependent magnetoresistance of the (Cu,C)-1223/LaMnO$_3$ film. It it clear that the superconductivity is gradually suppressed with increasing magnetic fields from 0 T to 9 T, and the superconducting transition becomes broader for $H\parallel c$ than for $H\parallel ab$, suggesting a stronger vortex motion in the case of $H\parallel c$ and the existence of anisotropy. In order to quantitatively calculate the anisotropy parameter $\Gamma$,  we extracted the upper critical field $H_{c2}$(T) for both field orientations by taking a criterion of 50\% the normal state resistivity $\rho _n$. Fig.~\ref{fig3}(c) displays the $H_{c2}$(T) of the film. It is obvious that the $H_{c2}(T)$ displays a linear  temperature dependence for the in-plane and out-of-plane magnetic fields. According to the simplified Werthamer–Helfand–Hohenberg (WHH) formula $H_{c2}(0)=-0.69\times (d H_{c2}/d T)|_{T_c}T_c$, we can  obtain the upper critical fields $H_{c2}^{ab}(0)=320$ T and $H_{c2}^{c}(0)=60$ T. The anisotropy parameters $\Gamma$ can be defined as $\Gamma=(m_c/m_{ab})^{1/2}=H_{c2}^{ab}/H_{c2}^{c}$. Thus the anisotropy parameter estimated from the ratio of $H_{c2}^{ab}(0)/H_{c2}^{c}(0)$ is about 5.3, which is very close to that of (Cu,C)-1234 single crystal\cite{HeSSaT2021}.

However,  the anisotropy parameter obtained by the above procedure may depend on the criterion of $H_{c2}$ definition. These is a more reliable way to determine the anisotropy at a certain temperature. According to the anisotropic Ginzburg-Landau theory, the angle dependent resistivity under different magnetic fields should be scaled to one curve by using the effective field $\tilde{H}=H\sqrt{cos^2\theta+\Gamma^{-2}sin^2\theta}$, where $\theta$ is the angle between the field and the c-axis. The scaling method has been successfully applied to different superconducting materials\cite{HeMTP2022,Jia2008,Liu2014}. The angle dependent resistance were measured under different magnetic fields as shown in Fig.~\ref{fig3}(d). And the data can be nicely scaled, yielding an anisotropy $\Gamma=5$ at 90 K. We can see that the anisotropy parameter determined from both ways is very close to each other. This result indicates that the (Cu,C)-1223 film grown on the Hastelloy tapes also has a low anisotropy.

\begin{figure*}
	\centering
	\includegraphics[width=0.5\textwidth]{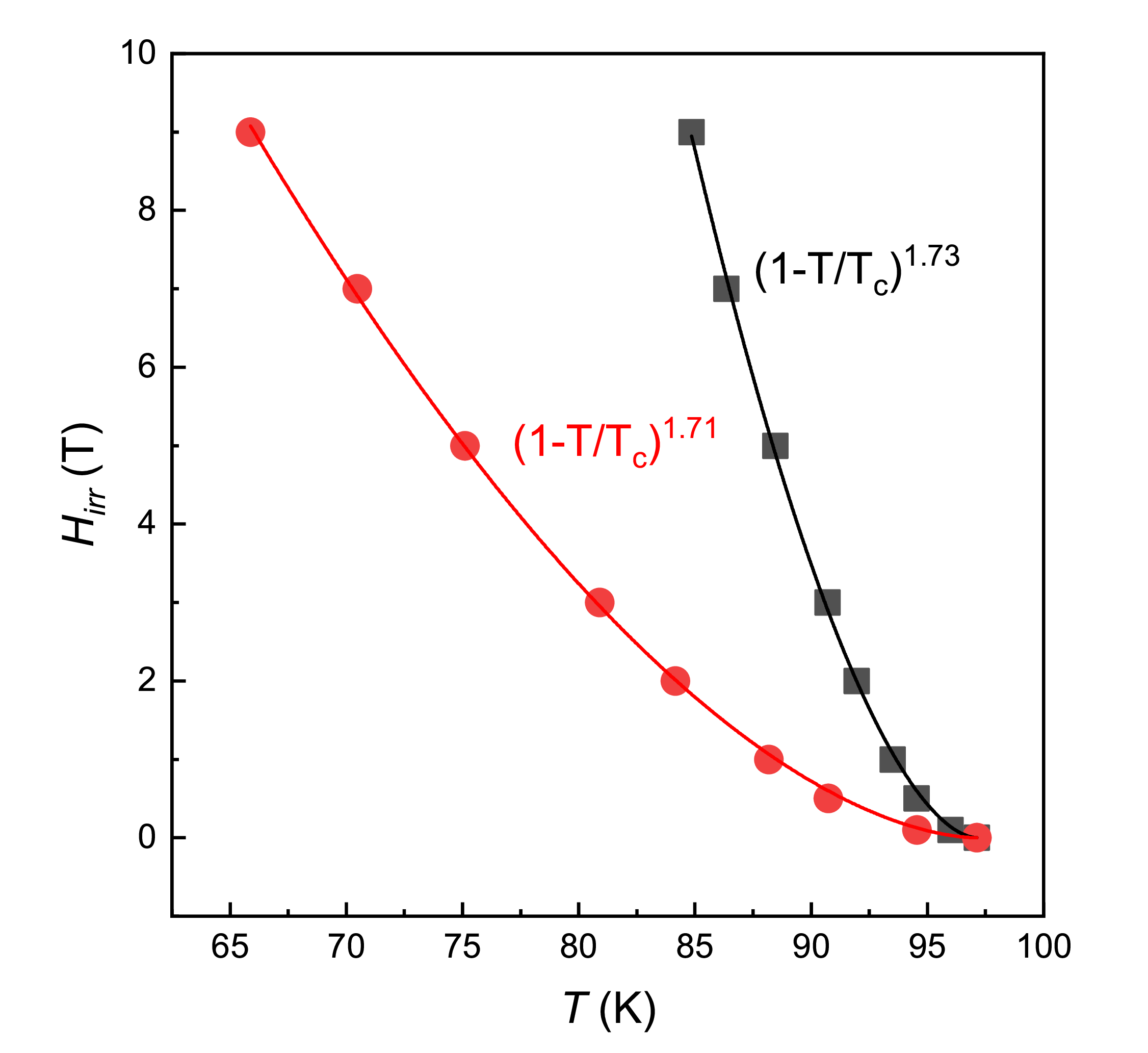}
	\caption{ Temperature dependence of the irreversibility field $H_{irr}$. The solid lines show the fitting results by $H_{irr}=H_{irr}(0) (1-T/T_c)^{\beta}$ with $\beta=1.71$ for $H\parallel c$ and $\beta=1.73$ for $H\parallel ab$.}
	\label{fig4}
\end{figure*}

In addition, the irreversibility fields can be determined by using a criterion of 1\% normal state resistivity $\rho _n$. Fig.~\ref{fig4} shows the irreversibility fields under the magnetic field $H\parallel c$ and $H\parallel ab$.  The data are nicely fitted with the power exponent
\begin{equation}
	H_{irr}=H_{irr}(0) (1-T/T_c)^{\beta}
\end{equation}
where $\beta$ =1.71 and 1.73 for $H\parallel c$ and $H\parallel ab$. These values are very close to (Cu,C)-1234 film\cite{DuanSSaT2020}. we can obtain the irreversibility field $H_{irr}=$ 4.2 T at 77 K by extrapolation.

\begin{figure*}
	\centering
	\includegraphics[width=\textwidth]{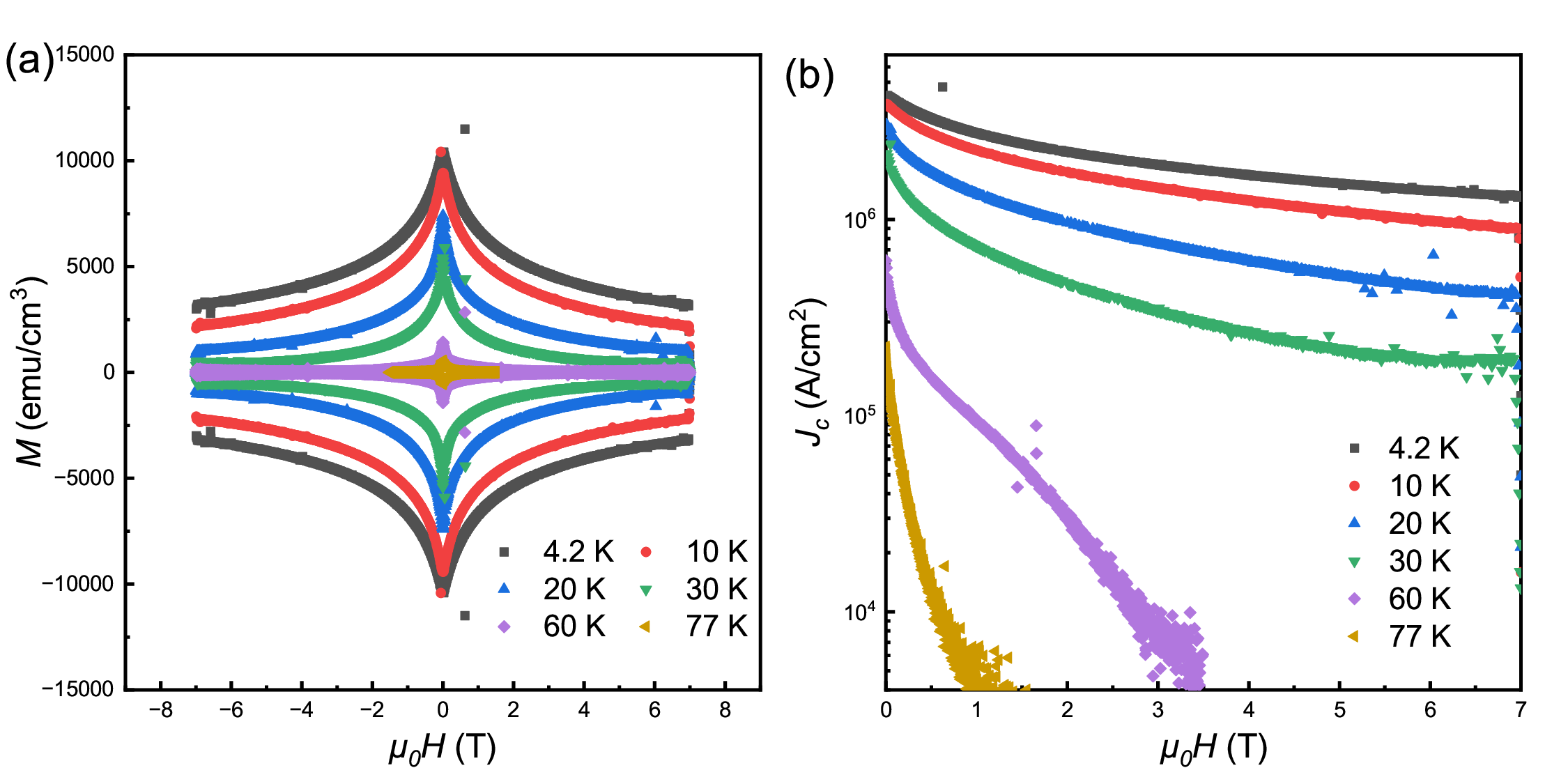}
	\caption{(a) The magnetization hysteresis loops after the paramagnetic background subtraction measured at different temperatures with $H\parallel c$. (b) The magnetic field dependence of the critical current densities calculated by the Bean critical state model. }
	\label{fig5}
\end{figure*}

In order to evaluate the current carrying ability of the (Cu,C)-1223 film, we measured the magnetization hysteresis loops (MHLs) of the (Cu,C)-1223/LaMnO$_3$ film under different temperature. Fig.~\ref{fig5}(a) displays the magnetization hysteresis loops after the paramagnetic background subtraction. We take the value of $M_{bg}=(M(H^+)+M(H^-))/2$ as the background signal from the substrate, and this background is removed in order to calculate the critical current density. Here $H^+$ and $H^-$ are the magnetic field in the ascending and descending processes. Since the background signal from the substrate has a soft-ferromagnetic feature without hysteresis, it is thus reasonable to derive the bulk critical current density following above mentioned treatment. By using the Bean critical state model, the critical current density $J_c$ can be calculated by the formula
\begin{equation}
J_c=20\Delta M/[w(1-w/3l)]
\end{equation}
Here $\Delta M$ is the width of the MHL; w and l are the width and the length of the sample, respectively. The calculated critical current density with varied magnetic fields is displayed in Fig.~\ref{fig5}(b). The calculated $J_c $(0 T) reaches $3.8 \times 10^6$ A/cm$^2$ at 10 K, which is one order of magnitude higher than the value of (Cu,C)-1234 films grown on single-crystal substrate \cite{DuanSSaT2020}.

Furthermore, the critical current density, anisotropy and irreversibility field of the (Cu,C)-1223 film in the liquid nitrogen boiling temperature are significantly higher than those of Bi-2223\cite{Clayton2004} and close to those of YBCO. The onset transition temperature is also higher than those of Bi-2223 and YBCO. Moreover,  the critical temperature of the (Cu,C)-1223 bulk sample can reach 120 K\cite{KodamaPCS2003} and the irreversible field can reach 9.4 T at 77 K\cite{IyoPB2000}. This indicates that the properties of the film can be enhanced by optimizing the growth conditions. Most importantly, our present work demonstrates that the (Cu,C)-1223 film can be grown on flexible metallic substrates using the same fabrication technique as the second generation coated conductor YBCO. This means that (Cu,C)-1223 coated conductors can be manufactured using the same industrial processes currently employed for YBCO to meet the requirements of industrial applications. Consequently, these results suggest that (Cu,C)-1223 may have significant potential for applications in the liquid nitrogen temperature region. For future applications of this material, its superconducting properties at low temperature and high fields are also crucial, which may be the next focus for further investigation.
\section{Conclusions}
The c-axis oriented (Cu,C)-1223 films have been successfully fabricated by pulsed laser deposition technique on flexible Hastelloy tapes with the top buffer layers of LaMnO$_3$ and CeO$_2$. The zero-resistance transition temperatures are 96 K and 98 K, respectively. The anisotropy, the irreversibility fields and the critical current densities of the (Cu,C)-1223/LaMnO$_3$ film are investigated, showing great potential for applications. The present work will promote fundamental research and practical applications of (Cu,C)-1223 in the liquid nitrogen temperature region, or for high field applications in the temperature region of 20 K. 

\section*{Data availability}
Data will be made available on a reasonable request.

\section*{Note for property right}
The property right is protected by the Chinese patent No. 202410414918.4.

\section*{Acknowledgements}
We appreciate the kind help in the early establishment of the PLD deposition equipment by Hai-Feng Chu and Tianfeng Duan. We are grateful for the useful discussions with Prof. David Larbalestier. The Hastelloy tapes with a LaMnO$_3$ and CeO$_2$ cap layer were kindly provided by Dr. Hongwei Gu in the company of Suzhou Advanced Materials, and Dr. Jiamin Zhu from the company of Shanghai Superconductor, respectively. The project is supported by the National Key Research and Development Program of China  (No. 2022YFA1403201), National Natural Science Foundation of China (Nos. 11927809, 12061131001).

\section*{References}
\providecommand{\noopsort}[1]{}\providecommand{\singleletter}[1]{#1}%


\begin{thebibliography}{10}
	
	\bibitem{Onnes1911}
	H.K. Onnes.
	\newblock The resistance of pure mercury at helium temperatures.
	\newblock {\em Commun. Phys. Lab. Univ. Leiden}, 12, 01, 1911.
	
	\bibitem{Zhang2019}
	Pingxiang Zhang, Jianfeng Li, Qiang Guo, Yanmin Zhu, Kaijuan Yan, Ruilong Wang,
	Kailin Zhang, Xianghong Liu, and Yong Feng.
	\newblock {\em {NbTi} superconducting wires and applications}, pages 279--296.
	\newblock Elsevier, 2019.
	
	\bibitem{Xu2017}
	Xingchen Xu.
	\newblock A review and prospects for {Nb$_3$Sn} superconductor development.
	\newblock {\em Superconductor Science and Technology}, 30(9):093001, 2017.
	
	\bibitem{SchillingN1993}
	A.~Schilling, M.~Cantoni, J.~D. Guo, and H.~R. Ott.
	\newblock Superconductivity above {130 K} in the {Hg–Ba–Ca–Cu–O}
	system.
	\newblock {\em Nature}, 363(6424):56--58, 1993.
	
	\bibitem{ShengN1988}
	Z.~Z. Sheng and A.~M. Hermann.
	\newblock Bulk superconductivity at {120 K} in the {Tl-Ca/Ba–Cu–O} system.
	\newblock {\em Nature}, 332(6160):138--139, 1988.
	
	\bibitem{MaedaJJoAP1988}
	Hiroshi Maeda, Yoshiaki Tanaka, Masao Fukutomi, and Toshihisa Asano.
	\newblock A new {high-T$_c$} oxide superconductor without a rare earth element.
	\newblock {\em Japanese Journal of Applied Physics}, 27(2A):L209, 1988.
	
	\bibitem{CavaNature1987}
	R.~J. Cava, B.~Batlogg, C.~H. Chen, E.~A. Rietman, S.~M. Zahurak, and
	D.~Werder.
	\newblock Oxygen stoichiometry, superconductivity and normal-state properties
	of {YBa$_2$Cu$_3$O$_7$}.
	\newblock {\em Nature}, 329(6138):423--425, 1987.
	
	\bibitem{Clayton2004}
	N~Clayton, N~Musolino, E~Giannini, V~Garnier, and R~Flükiger.
	\newblock Growth and superconducting properties of
	{Bi$_2$Sr$_2$Ca$_2$Cu$_3$O$_{10}$} single crystals.
	\newblock {\em Superconductor Science and Technology}, 17(9):S563--S567, 2004.
	
	\bibitem{AlNaafa2010}
	Mohammad Al-Naafa.
	\newblock New insights on the intrinsic anisotropy of {YBCO} films.
	\newblock {\em Journal of Superconductivity and Novel Magnetism},
	23(7):1407--1417, 2010.
	
	\bibitem{MacManusDriscoll2021}
	Judith~L. MacManus-Driscoll and Stuart~C. Wimbush.
	\newblock Processing and application of high-temperature superconducting coated
	conductors.
	\newblock {\em Nature Reviews Materials}, 6(7):587--604, 2021.
	
	\bibitem{KawashimaPCS1994}
	T.~Kawashima, Y.~Matsui, and E.~Takayama-Muromachi.
	\newblock New oxycarbonate superconductors
	{(Cu$_{0.5}$C$_{0.5}$)Ba$_2$Ca$_{n-1}$Cu$_n$O$_{2n+3}$ (n=3, 4)} prepared at
	high pressure.
	\newblock {\em Physica C: Superconductivity}, 224(1-2):69--74, 1994.
	
	\bibitem{KumakuraPCS1994}
	H.~Kumakura, K.~Togano, T.~Kawashima, and E.~Takayama-Muromachi.
	\newblock Critical current densities and irreversibility lines of new
	oxycarbonate superconductors
	{(Cu$_{0.5}$C$_{0.5}$)Ba$_2$Ca$_{n-1}$Cu$_n$O$_{2n+3}$ (n = 3, 4)}.
	\newblock {\em Physica C: Superconductivity}, 226(3-4):222--226, 1994.
	
	\bibitem{ZhangSA2018}
	Yue Zhang, Wenhao Liu, Xiyu Zhu, Haonan Zhao, Zheng Hu, Chengping He, and
	Hai-Hu Wen.
	\newblock Unprecedented high irreversibility line in the nontoxic cuprate
	superconductor {(Cu,C)Ba$_2$Ca$_3$Cu$_4$O$_{11}$}.
	\newblock {\em Science Advances}, 4(9), 2018.
	
	\bibitem{HeSSaT2021}
	Chengping He, Xue Ming, Jin Si, Xiyu Zhu, Jinhua Wang, and Hai-Hu Wen.
	\newblock Characterization of the {(Cu,C)Ba$_{2}$Ca$_{3}$Cu$_4$O$_{11+\delta}$}
	single crystals grown under high pressure.
	\newblock {\em Superconductor Science and Technology}, 35(2):025004, 2021.
	
	\bibitem{HeMTP2022}
	Chengping He, Xue Ming, Renju Lin, Xinwei Fan, Dongsheng Song, Binghui Ge, and
	Hai-Hu Wen.
	\newblock Key factor for low anisotropy and high irreversibility field in
	{(Cu,C)Ba$_2$Ca$_3$Cu$_4$O$_{11+\delta}$}.
	\newblock {\em Materials Today Physics}, 29:100913, 2022.
	
	\bibitem{DuanSSaT2020}
	Tianfeng Duan, Jiahao Hao, Haifeng Chu, and Hai-Hu Wen.
	\newblock Preparation and superconducting properties of the
	{(Cu,C)Ba$_{2}$Ca$_{3}$Cu$_4$O$_y$} films with zero-resistance transition
	temperature of {96 K}.
	\newblock {\em Superconductor Science and Technology}, 33(2):025009, 2020.
	
	\bibitem{DuanPCSaiA2020}
	Tianfeng Duan, Jiahao Hao, Haifeng Chu, Bowen Li, Yaomin Dai, and Hai-Hu Wen.
	\newblock Existence of carbonate clusters and its relationship with critical
	temperature in superconducting {(Cu,C)Ba$_{2}$Ca$_{3}$Cu$_4$O$_y$} films.
	\newblock {\em Physica C: Superconductivity and its Applications}, 573:1353646,
	2020.
	
	\bibitem{Ou2024}
	Meng-Jun Ou, Yuecong Liu, Yi~Wang, and Hai-Hu Wen.
	\newblock Superconducting thin films of {(Cu,C)Ba$_2$Ca$_2$Cu$_3$O$_{9+\delta}$}
	with zero resistance transition temperature close to {100 K}.
	\newblock Arxiv: 2408.02094.
	
	\bibitem{ChailloutPCS1996}
	C.~Chaillout, S.~Le~Floch, E.~Gautier, P.~Bordet, C.~Acha, Y.~Feng, A.~Suloice,
	J.L. Tholence, and M.~Marezio.
	\newblock Enhancement of tc of {C$_y$Cu$_{1-y}$Ba$_2$Ca$_2$Cu$_3$O$_x$} from
	{67 K} to {120 K} by reduction treatments.
	\newblock {\em Physica C: Superconductivity}, 266(3–4):215--222, 1996.
	
	\bibitem{Jia2008}
	Ying Jia, Peng Cheng, Lei Fang, Huan Yang, Cong Ren, Lei Shan, Chang-Zhi Gu,
	and Hai-Hu Wen.
	\newblock Angular dependence of resistivity in the superconducting state of
	{NdFeAsO$_{0.82}$F$_{0.18}$} single crystals.
	\newblock {\em Superconductor Science and Technology}, 21(10):105018, 2008.
	
	\bibitem{Liu2014}
	Jianzhong Liu, Delong Fang, Zhenyu Wang, Jie Xing, Zengyi Du, Sheng Li, Xiyu
	Zhu, Huan Yang, and Hai-Hu Wen.
	\newblock Giant superconducting fluctuation and anomalous semiconducting normal
	state in {NdO$_{1-x}$ F$_x$Bi$_{1-y}$S$_2$} single crystals.
	\newblock {\em EPL (Europhysics Letters)}, 106(6):67002, 2014.
	
	\bibitem{KodamaPCS2003}
	Y.~Kodama, A.~Iyo, M.~Hirai, H.~Kito, and Y.~Tanaka.
	\newblock Annealing study of superconducting properties in a {Cu-1223}
	superconductor using {O$_2$-HIP} apparatus.
	\newblock {\em Physica C: Superconductivity}, 392–396:77--81, 2003.
	
	\bibitem{IyoPB2000}
	Akira Iyo, Yasumoto Tanaka, Norio Terada, Madoka Tokumoto, and Hideo Ihara.
	\newblock Annealing effect on the irreversibility line in {(Cu, C)Ba$_2$ Ca$_2$
		Cu$_3$ O$_y$ }.
	\newblock {\em Physica B}, 284-288(Japan):867--868, 2000.
	
\end{thebibliography}
\end{document}